# A Resource Allocation Mechanism for Video Mixing as a Cloud Computing Service in Multimedia Conferencing Applications

Abbas Soltanian[†], Mohammad A. Salahuddin[†‡], Halima Elbiaze[‡], Roch Glitho[†]
[†]Concordia University, Montreal, Quebec, Canada, [‡]Université du Québec À Montréal, Montreal, Quebec, Canada
ab_solta@encs.concordia.ca, mohammad.salahuddin@ieee.org, elbiaze.halima@uqam.ca, glitho@ciise.concordia.ca

*Abstract*—Multimedia conferencing is the conversational exchange of multimedia content between multiple parties. It has a wide range of applications (e.g. Massively Multiplayer Online Games (MMOGs) and distance learning). Many multimedia conferencing applications use video extensively, thus video mixing in conferencing settings is of critical importance. Cloud computing is a technology that can solve the scalability issue in multimedia conferencing, while bringing other benefits, such as, elasticity, efficient use of resources, rapid development, and introduction of new applications. However, proposed cloud-based multimedia conferencing approaches so far have several deficiencies when it comes to efficient resource usage while meeting Quality of Service (QoS) requirements. We propose a solution to optimize resource allocation for cloud-based video mixing service in multimedia conferencing applications, which can support scalability in terms of number of users, while guaranteeing QoS. We formulate the resource allocation problem mathematically as an Integer Linear Programming (ILP) problem and design a heuristic for it. Simulation results show that our resource allocation model can support more participants compared to the state-of-the-art, while honoring QoS, with respect to end-to-end delay.

## I. INTRODUCTION

Multimedia conferencing[1] can be defined as conversational and real time exchange of multimedia content, such as, audio and video, between several parties [1]. It has three main architectural components, namely signaling, media handling, and conference control [2]. Video mixing is a functionality provided by the media handling component. It mixes different video sources from conferencing participants to reach one video stream as output.

Several conferencing applications exist, which use video extensively, such as, distance learning, video conferencing, and Massively Multiplayer Online Games (MMOGs). Therefore, video mixing is of critical importance in conferencing applications. There might be thousands or hundreds of thousands of participants scattered over large geographical areas in some conferencing applications like MMOGs [3], thus requiring scalability. Furthermore, the pressure of cost reduction brings about the need for efficient use of resources.

Cloud computing is an emerging paradigm for provisioning computing and storage infrastructure and services with three key facets: Software as a Service (SaaS), Platform as a Service (PaaS), and Infrastructure as a Service (IaaS) [3]. It has several inherent benefits, such as, efficient use of resources, scalability and elasticity. These characteristics make it suitable for provisioning of conferencing applications.

This paper deals with video mixing as a SaaS for conferencing applications. Fig. 1 depicts the assumed business model, where conferencing applications are offered as services to end-users. These applications rely on a conferencing service that is also offered as a SaaS. Video mixing as a service, offered to conferencing service providers, relies on geographical distributed IaaS, providing the actual resources (e.g. CPU, RAM, storage) needed for video mixing. The key component of IaaSs is in red in Fig. 1, which is the focus of paper, that is, the video mixing resource allocator (VMRA).

VMRA is a dynamic resource allocation mechanism, since the demand for video mixing depends not only on the number of participants, but also on how the participants use the video resource. Furthermore, it caters to QoS, with respect to video mixing response time. It performs a fine-grained resource and virtual machine (VM) scaling, to improve efficiency in resource utilization, while meeting the QoS requirements of video mixing service in conferencing applications.

We analyze our proposed resource allocation mechanism theoretically by modeling it as an optimization problem. Moreover, we design a heuristic for real-world scenarios. The results show that our mechanism outperforms current state-of-the-art in maximizing resource utilization, while meeting QoS, across multiple IaaSs. In addition, compared to the state-of-the-art, in our model, a video mixer can accommodate higher numbers of participants without sacrificing QoS.

## II. REQUIREMENTS AND RELATED WORK

### A. Requirements

A crucial requirement for video mixing as a service pertains to dynamic scalability, or accommodating the changing number of participants. For example, in one study, the number of users in World of Warcraft (WoW), fluctuates between 1.5 and 2.5 million over 10 hours [4]. Therefore, video mixing resource allocator should be able to dynamically scale required resources. Efficient use of resources is another important requirement. As an example, WoW uses more than ten-thousand servers, while most of the servers' capacities remain idle most of the time [4].

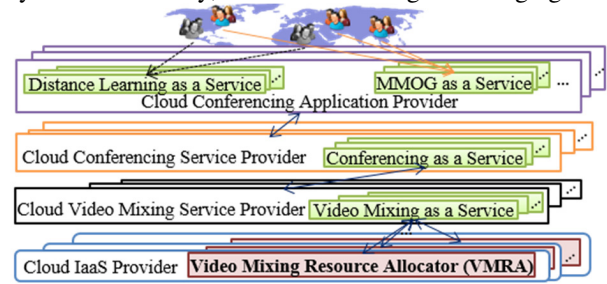

Fig. 1. Business model

---
[1] Here onwards referred to as *conferencing*.

Moreover, meeting QoS requirements, such as jitter, throughput, and end-to-end delay, is crucial in video mixing as a service. In our study, we focus on end-to-end delay. Based on International Telecommunication Union (ITU), total end-to-end delay in conferencing should not exceed 400 msec [5]. Video mixing response time is a critical component of the total end-to-end delay in conferencing. It can be defined as the time between arrival of video mixing request and departure of video mixing result back to the user. Video mixing resource allocator should consider it in order to appropriately provision resources. In addition, it brings the need of relying on video mixing as a service from multiple geographically distributed IaaSs.

*B. Related Work*

*1) Traditional Resource Allocation for Video Mixing*

Most existing resource allocation solutions have been proposed for peer-to-peer (P2P) conferencing and centralized multimedia conferencing [6]. Yuen and Chan [7] attempt to reduce worst-case video transmission delay from different video sources to users. They propose an algorithm to select peers as mixers to achieve minimum overall delay. However, their algorithm does not account for video mixing response time. Chen *et al.* [8] also propose P2P multi-party video conferencing solution to achieve low end-to-end delay. They optimize the streaming rates of all the peers subject to network bandwidth constraints. Their study reduces end-to-end delay without tackling the specifics of video mixing.

Multipoint Control Unit (MCU) [9] is a media handling component that could include video mixing as a functionality. Traditionally, all video mixing requests are handled by a single MCU, where resources are allocated in a static manner. Thus, this approach is not scalable and uses resources inefficiently.

*2) Resource Allocation for Video Mixing in Cloud*

Liao *et al.* [10] focus on minimizing video transmission delay and consequently the total end-to-end delay. Their heuristic reduces average and maximum end-to-end delay by choosing a network of servers and clients as mixers, reducing the delay between conference endpoints. However, since they allocate all available resources to the mixer, their resource allocation scheme does not meet the efficient resource usage requirement. Zhang *et al.* [6] propose minimizing mean end-to-end delay by choosing the best physical servers . They find the ideal geographic server locations and map to the closest physical server candidates. Efficient resource usage is a limitation of their work, as they allocate the entire server resources as a mixer. Moreover, they do not consider video mixing response time in their end-to-end delay.

Taheri *et al.* [1] propose a cloud infrastructure that relies on conferencing substrates. Their architecture enables different conferencing applications to be built using virtualized conferencing substrates that can be provided by different substrate providers. Li *et al.* [11] offer conferencing as a cloud-based service. They follow the structure of Service Oriented Architecture (SOA) to propose a design for cloud-based conferencing. However, none of these work tackle video mixing resource allocation.

*3) Other Approaches*

Negralo, *et al.* [12] and Weng and Wang [13] have addressed the resource allocation problem for conferencing applications. Others ([14], [15], [16]) focus more on optimizing resource allocation to reduce cost. Nan *et al.* [14] used a queuing model to optimize resource allocation. They studied VM allocation problem for multimedia application providers and minimized the resource cost under the end-to-end delay requirement [15]. Sembiring and Beyer [16] propose a dynamic cloud resource allocation to different multimedia tasks with respect to system efficiency and QoS. However, none of these literatures rely on fine-grained resource scaling. Moreover, they do not tackle the specifics of video mixing.

Compared to the state-of-the-art, our work fills the need of a resource allocation scheme for cloud-based conferencing applications that (i) considers the specifics of video mixing as a service, (ii) meets the QoS requirements, and (iii) scales dynamically, while using resources in an efficient way.

III. SYSTEM MODEL

Our system model includes cooperation, video mixing, and mathematical models. In our mathematical model, we define VMRA as an Integer Linear Programming (ILP) problem.

*A. Cooperation Model*

We consider a large-scale distributed cloud infrastructure to support conferencing applications and video mixing as a service, consisting of users, separate zones and an IaaS in each zone $z$, as depicted in Fig. 2. We illustrate users scattered across a large geographical area, wanting to join a conferencing application, such as MMOG. We assume that in each zone $z$, there is a data center providing IaaS, where each data center consists of a number of servers ($N_z$), hosting VMs. Furthermore, we assume that zones are interconnected in a full mesh manner. The same assumption applies to VMs in a data center, as shown in Fig. 2.

Users in each zone will connect to their local data center to join a conferencing application. Each user is considered as a video source, sending video and requesting video mixing service. The challenge lies in allocating the resources for video mixing to achieve optimal resource utilization, while guaranteeing QoS requirements.

*B. Video Mixing Model*

VMRA decides to add resources to existing VMs or create a new VM when a video source is added to a data center. Adding resources is done in fine granularity. This implies that VMRA will add minimal required resources in an elastic manner. It will also balance the load between all the VMs in a data center. After provisioning appropriate resources, a video source will join a VM, that is, a video mixer and video mixing will start. The video mixing process is illustrated in Fig. 3.

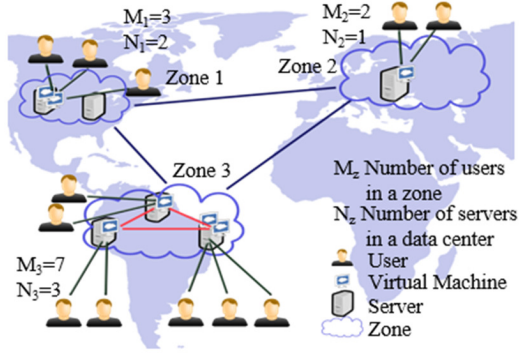

Fig. 2. Communication model b/w VMs in a data center and across zones

Our video mixing model follows the Fork/Join parallelism technique [17]. All video mixing requests in a data center fork off to several other mixing processes, which are concurrently executed in each VM, until they finally join into a single mixed video. VMs mix their video sources in parallel. Therefore, the required time for this step depends on the maximum number of video sources connected to any VM ($V_z$) in zone $z$.

Each VM will send the result to other VMs in the same data center. This intra-zone video exchange time is in $T_{int}$. Next, each VM mixes the incoming videos from other VMs with the result of its own mixed video source. The time for this step depends on the total number of VMs in the data center.

Here, mixed video of a data center is ready and sent to all other data centers. This inter-zone video exchange time is in $T_{ext}$. Then, VMs will start mixing the incoming videos from other zones with the one of their own zone. Here, the required time depends on the total number of zones and the mixed video across all zones is ready to be sent back to the users.

### C. Mathematical Model

This subsection, presents our VMRA problem formulation, which is modeled as an ILP problem.

*1) Problem Statement*

Given a data center with $N_z$ servers and $M_z$ users (video sources), let $T_{mix(k)}$ and $R_{mix(k)}$ represent the time and the resource required to mix $k$ video sources, respectively. Also, let $T_{int}$ and $T_{ext}$ denote the time to exchange a video across VMs and zones, respectively. $R_O$ are the resources dedicated to VM operation, hence, they cannot be utilized for video mixing. There are thresholds $T_\varepsilon$ on QoS, pertaining to the maximum acceptable video mixing response time, and $R_\varepsilon$ on server resource capacity, respectively. Find the minimum number of VMs, while efficiently using resources and respecting QoS.

We model this as an ILP problem, where we assume a video mixer to be analogous to a VM. Tables I and II delineate the inputs and variables of our problem, respectively.

*2) Objectives*

We assume the operational cost of a VM, in terms of non-utilizable resources, supersedes the cost of resources required for handling the video mixing request of a participant, as in (1). Furthermore, we assume homogeneous costs of video mixing resources across servers. Therefore, the operational cost $R_O$, associated with a VM, inhibits the introduction of a new VM, in the event of a new participant arrival. That is, a new VM is only instantiated if an incoming request cannot be handled by increasing the resource of an existing VM.

$$R_O \gg (R_{mix\,(k+1)} - R_{mix(k)}) \quad (1)$$

Equation (2) depicts our multiple objectives. Primarily, we minimize the allocated resources across all zones, by minimizing the number of VMs. On the other hand, the time to mix videos in zone $z$ depends on the maximum number of users connected to a VM ($V_z$). We balance the load between VMs to decrease the overall video mixing time. Note that these are competing objectives. Therefore, we prioritize minimizing the number of VMs by normalizing $V_z$ with the maximum number of users in zone $z$.

$$minimize \left\{ \sum_{i=1}^{N_z} \sum_{j=1}^{M_z} x_{i,j} + \frac{V_z}{M_z} \right\} \quad (2)$$

*3) Constraints*

VMs and users cannot be split across multiple servers and VMs, respectively. Equation (3) ensures that a VM exists on a single server. Similarly, (4) allows a user to connect to a single VM. Furthermore, if there are users connected to a VM, that VM should exist on one server, as depicted in (5) and (6).

$$\sum_{i=1}^{N_z} x_{i,j} \leq 1 \qquad \forall 1 \leq j \leq M_z \quad (3)$$

$$\sum_{j=1}^{M_z} y_{j,k} = 1 \qquad \forall 1 \leq k \leq M_z \quad (4)$$

$$\sum_{k=1}^{M_z} y_{j,k} \leq \beta \cdot \left( \sum_{i=1}^{N_z} x_{i,j} \right) \qquad \forall 1 \leq j \leq M_z \quad (5)$$

$$\sum_{k=1}^{M_z} y_{j,k} \geq \sum_{i=1}^{N_z} x_{i,j} \qquad \forall 1 \leq j \leq M_z \quad (6)$$

Video mixing required resources, that is, the VMs operating resources and their connected number of users, is bounded by the server resource capacity $R_\varepsilon$, in (7).

$$\quad (7)$$

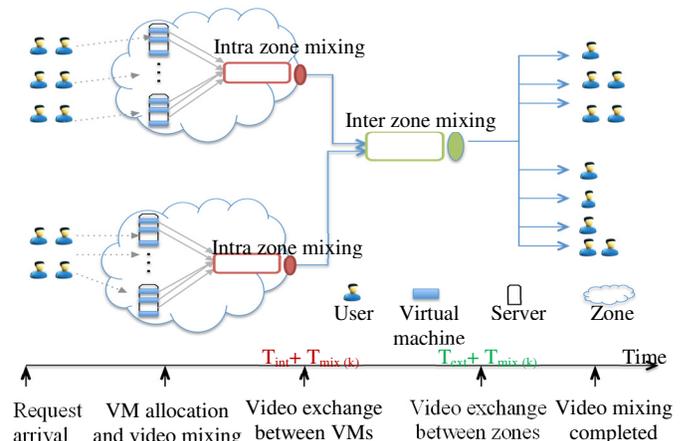

Fig. 3. An example of our video mixing model

TABLE I. Problem inputs

| Input | Definition |
|---|---|
| $Z$ | number of zones |
| $N_z$ | number of servers in zone $z$ |
| $M_z$ | number of users i.e. video sources in zone $z$ |
| $T_{int}$ | time to send a video between VMs in a zone |
| $T_{ext}$ | time to send a video between zones, $Z = 1 \Rightarrow T_{ext} = 0$ |
| $T_{mix(k)}$ | time to mix $k$ video sources, $T_{mix(1)} = 0$ |
| $T_\varepsilon$ | QoS threshold (acceptable mixing response time) |
| $R_{mix(k)}$ | required resources for mixing $k$ video sources in a VM |
| $R_O$ | non-utilizable VM operating resources |
| $R_\varepsilon$ | threshold on the maximum amount of resources on a server |
| $\beta$ | large enough constant |

TABLE II. Problem variables

| Variable | Definition | |
|---|---|---|
| $X$ | $N_z \times M_z$ binary matrix, where | $x_{i,j} = \begin{cases} 1, & \text{if server } i \text{ hosts VM } j \\ 0, & \text{otherwise} \end{cases}$ |
| $Y$ | $M_z \times M_z$ binary matrix, where | $y_{j,k} = \begin{cases} 1, & \text{if user } k \text{ is connected to VM } j \\ 0, & \text{otherwise} \end{cases}$ |
| $V_z$ | Maximum number of users that are connected to a VM in zone $z$ | |
| $U$ | A vector where $u_j$ is the number of users connected to VM $j$ | |
| $C$ | $N_z \times M_z$ matrix, where | $c_{i,j} = \begin{cases} u_j, & \text{if server } i \text{ hosts VM } j \\ 0, & \text{otherwise} \end{cases}$ |

Note that the product $\sum_{j=1}^{M_z}(x_{i,j} \cdot \sum_{k=1}^{M_z} y_{j,k})$ in (7) is non-linear. Therefore, we linearize (7) by replacing it with constraints (8)-(13).

$$\sum_{k=1}^{M_z} y_{j,k} = u_j \quad \forall 1 \leq j \leq M_z \quad (8)$$

$$c_{i,j} \leq M_z \cdot x_{i,j} \quad \forall 1 \leq i \leq N_z, \forall 1 \leq j \leq M_z \quad (9)$$

$$c_{i,j} \leq u_j \quad \forall 1 \leq i \leq N_z, \forall 1 \leq j \leq M_z \quad (10)$$

$$c_{i,j} \geq u_j - M_z(1 - x_{i,j}) \quad \forall 1 \leq i \leq N_z, \forall 1 \leq j \leq M_z \quad (11)$$

$$c_{i,j} \geq 0 \quad \forall 1 \leq i \leq N_z, \forall 1 \leq j \leq M_z \quad (12)$$

$$R_O \cdot \left(\sum_{j=1}^{M_z} x_{i,j}\right) + R_{mix\left(\sum_{j=1}^{M_z} c_{i,j}\right)} \leq R_\varepsilon \quad \forall 1 \leq i \leq N_z \quad (13)$$

The maximum number of users, $V_z$, in a zone $z$ influences the video mixing time. Equation (14) finds $V_z$, for each zone.

$$\sum_{k=1}^{M_z} y_{j,k} \leq V_z \quad \forall 1 \leq j \leq M_z \quad (14)$$

Video mixing response time for a zone $z$, depends on the maximum number of users connected to a single VM in that zone ($T_{mix(V_z)}$). Note that VMs should mix the output of video mixing from other VMs too, therefore, the video mixing response time will also be influenced by the total amount of VMs across all servers in $z$. This time is given by $T_{mix\left(\sum_{i=1}^{N_z} \sum_{j=1}^{M_z} x_{i,j}\right)}$, with an inter-zone exchange time of $T_{int}$. Furthermore, VMs should mix the incoming videos from all other zones, time for which is represented by $T_{mix(Z)}$, with an intra-zone exchange time of $T_{ext}$. Equation (15) ensures that this total video mixing response time for each zone $z$, abides by the QoS threshold $T_\varepsilon$.

$$T_{mix(V_z)} + T_{int} + T_{mix\left(\sum_{i=1}^{N_z} \sum_{j=1}^{M_z} x_{i,j}\right)} + T_{ext} + T_{mix(Z)} \leq T_\varepsilon$$

$$\forall 1 \leq z \leq Z \quad (15)$$

VMRA executes in each zone separately. However, because video mixing as a service relies on multiple IaaSs, the total number of zones will influence VMRA's decision. Based on (15), different response times across zones are attributed to the different values of $T_{mix(V_z)}$ and $T_{mix\left(\sum_{i=1}^{N_z} \sum_{j=1}^{M_z} x_{i,j}\right)}$. Zone $z$ will send its mixed video to other zones and wait to receive from them. Waiting time in (16) will add to the video mixing response time of zones that perform video mixing faster than the other zones. Thus, the video mixing response time will be equal to the maximum response time across all zones.

$$\begin{cases} MAX \left\{ \begin{array}{l} \left(T_{mix(V_p)} + T_{mix\left(\sum_{i=1}^{N_p} \sum_{j=1}^{M_p} x_{i,j}\right)}\right) - \\ \left(T_{mix(V_z)} + T_{mix\left(\sum_{i=1}^{N_z} \sum_{j=1}^{M_z} x_{i,j}\right)}\right) \end{array} \right\} & \forall 1 \leq p \leq Z \\ 0, \text{if } MAX \leq 0 \end{cases} \quad (16)$$

## IV. VMRA Heuristic

Based on (1), VMRA always processes a new mixing request by adding required resources to the existing VMs unless it cannot satisfy the QoS requirement or there are not enough free resources on the server. In this case VMRA instantiates a new VM and balances the load between VMs in the data center. Load balancing helps minimizing the maximum number of connected users to each VM. We achieve this by employing MinMax our objective, that is, the minimization of the maximum number of users on VMs and consequently, based on (15), it decreases the total response time.

VMRA checks the available resources when it decides to instantiate a new VM. Moreover, it checks the possibility of satisfying QoS requirement, by adding a new VM. Our heuristic is as described in Algorithm 1. We consider the constants and variables shown in Table I and Table II as the input to this algorithm.

**Algorithm 1.** Video mixing resource allocation

**Input:**
$Max\_M$ = M; // Max number of users that can be served in DC
$\alpha = 0$; // number of VMs
$\beta = 1$; // number of used servers
$R_\beta = R_\varepsilon$; // available resources on server $\beta$
Remain_User = 0; // auxiliary variable to scatter users between VMs
**Output:** $\alpha, U, Max\_M$

1.  **For each** m ∈ M **do**

*Phase 1: Test if there is a VM with lower users than $V_z$*

2.    **If** ($R_\beta \geq R_{mix(1)}$) **Then**
3.      **For** j = 1 → $\alpha$ **do**
4.        **If** ($u_j < V_z$) **Then**
5.          $u_j \leftarrow u_j + 1$
6.          Break, serve next **m**
7.      **end for**
8.    **end if**

*Phase 2: Create first VM in DC*

9.    **If** ($\alpha == 0$) **Then**
10.     $\alpha \leftarrow 1$
11.     $u_1 \leftarrow 1$
12.     $V_z \leftarrow 1$
13.   **end if**

*Phase 3: Test response time by increasing $V_z$ without adding VM*

14.   **Else if** ($R_\beta \geq R_{mix(1)}$ **AND** Response time($V_z \leftarrow V_z + 1, \alpha) \leq T\varepsilon$)**Then**
15.     $u_1 \leftarrow u_1 + 1$
16.     $V_z \leftarrow V_z + 1$
17.   **end else if**

*Phase 4: Test response time by adding a new VM on the same server*

18.   **Else if** ($R_\beta \geq R_{mix(1)} + R_O$) **Then**
19.     **If** (Response time($V_z \leftarrow \left\lceil \frac{m}{\alpha+1} \right\rceil, \alpha \leftarrow \alpha + 1) \leq T\varepsilon$) **Then**

| | |
|---|---|
| 20. | $\alpha \leftarrow a + 1$ |
| 21. | Remain_User ← m |
| 22. | **For** j = $\bar{\alpha} \to 1$ **do** |
| 23. | $u_j \leftarrow$ Remain_User / j |
| 24. | Remain_User ← Remain_User $- u_j$ |
| 25. | **end for** |
| 26. | $V_z \leftarrow \left\lceil \frac{m}{\alpha} \right\rceil$ |
| 27. | **end if** |
| 28. | **Else** |
| 29. | $Max_M \leftarrow m - 1$ |
| 30. | Break, DC cannot serve **m** users |
| 31. | **end else** |
| 32. | **end else if** |
| *Phase 5: Test response time by adding new VM on the other server* | |
| 33. | **Else If** ( ($N_z - \beta > 0$) **AND** ($R_\beta \geq R_{mix(1)} + R_O$)) **Then** |
| 34. | **If** (Response time($V_z \leftarrow \left\lceil \frac{m}{\alpha+1} \right\rceil$, $\alpha \leftarrow a + 1) \leq T_\varepsilon$) **Then** |
| 35. | $\beta \leftarrow \beta + 1$ |
| 36. | $\alpha \leftarrow a + 1$ |
| 37. | Remain_User← m |
| 38. | **For** j = $\bar{\alpha} \to 1$ **do** |
| 39. | $u_j \leftarrow$ Remain_User / j |
| 40. | Remain_User ← Remain_User $- u_j$ |
| 41. | **end for** |
| 42. | $V_z \leftarrow \left\lceil \frac{m}{\alpha} \right\rceil$ |
| 43. | **end if** |
| 44. | **Else** |
| 45. | $Max_M \leftarrow m - 1$ |
| 46. | Break, DC cannot serve **m** users |
| 47. | **end else** |
| 48. | **end else if** |
| 49. | **Else** |
| 50. | $Max_M \leftarrow m - 1$ |
| 51. | Break, DC cannot serve **m** users |
| 52. | **end for each** |
| **Return** $\alpha, U, Max\_M$ | |

In phase 1, VMRA tries to find a VM with lowest number of connected users. If VMRA finds such a VM, it will add required resources to that VM and assigns the new user to it. In phase 2, the first user wants to join. So, VMRA will create the first VM and assign that user to it. VMRA will reach phase 3 if all the VMs have the same number of users. Here, VMRA checks the available resources and the feasibility of satisfying QoS requirements, if it assigns a new user to one of the existing VMs. This assignment is crucial as it increases $V_z$, thus, impacting the video mixing time.

If increasing $V_z$ causes sacrificing QoS, VMRA decides to instantiate a new VM on the same server or on other servers based on available resources, in phase 4 and 5, respectively. If there are available resources, but VMRA cannot find any feasible solution to satisfy QoS, it will stop accepting new users in both phases 4 and 5.

This algorithm has a nested loop and its time complexity is based on the number of iterations of each loop. Therefore, the time complexity of our VMRA algorithm is $O(M_z \cdot \alpha)$.

## V. SIMULATION RESULTS

### A. Comparison Baselines

We compare VMRA with (i) popular traditional MCU [9], for video mixing, (ii) Nan *et al.* [14], cost minimization model in cloud, for a single class service, and (iii) cloud-based MCU (CMCU), which avoids upfront resource costs. However, since these models do not support multi-zone video mixing, we assume that each model is implemented in a zone and exchange mixed video amongst each other, until all sources are mixed.

### B. Environment and Settings

We assume a MMOG, where player's video is shared in the logic of the game and developed a custom simulator in JAVA. We simulate multiple data centers and game players as conferencing participants. VMRA heuristic runs on each data center part in our simulator. Players send their video mixing requests to the local data center and receive the result from it. Total number of game players across all zones fluctuates, since they can join or leave the game whenever they want to. For our simulation, we assume a snap-shot of the number of players in each zone. Our simulation parameters are depicted in Table III.

### C. Results

We simulate our heuristic to check supported number of users, resource utilization and video mixing response time.

#### 1) Number of Users

It is evident from Fig. 4, that VMRA can serve more users in a single zone in comparison to other baselines. This is because VMRA has the leverage to increase resources whenever it reaches the QoS threshold in contrast to the queuing model, where the number of computation nodes is fixed. VMRA also performs better than MCU and CMCU. Due to their centralized nature, both MCU and CMCU models leverage a single server entity and consequently are not equipped to handle large number of users.

When we increase the number of zones, we have to account for the inter-zone communication time of mixing videos. As a result, to satisfy video mixing response time threshold, video mixing as a service can serve a lower number of users in each zone, while the number of zones increase. Although there is a tradeoff between the number of zones and the number of users that can be served in each zone, total number of users that can be served across all zones will increase, as depicted in Fig. 5. In addition, VMRA shows a better growth rate, thus it shows better scalability, in terms of the number of users, in comparison to the other models.

#### 2) Resource Utilization and Video Mixing Response Time

Required resources for video mixing depends on the maximum number of served users. Accordingly, we study two different scenarios, each with a different number of video mixing requests: (i) *Meet-By-All* - In this scenario, we assume that there exists a maximum number of users, which can be served by all the resource allocation models in a zone, while respecting QoS. (ii) *Meet-By-Some* - In this scenario, we assume for all models, the number of users to be the maximum supported by VMRA, while respecting QoS. In this scenario, we relax the QoS constraint for the other models, giving them the leverage to support a higher number of users.

##### a) Resource Utilization: Meet-By-All Scenario

Fig. 6(a) and 6(b), depict the average and the maximum allocated resources over the total available resources in a data center, respectively. In MCU, because of upfront resource over provisioning, there are always some idle resources, which remain unutilized. However, because the allocated resources in MCU are always at 100%, we do not show it in the resource

TABLE III. Simulation parameters

| Parameter | Value | Parameter | Value | Parameter | Value |
|---|---|---|---|---|---|
| Z | 1-6 | $T_{mix(k)}$ | 7 msec | $R_{mix(k)}$ | 20 MB (RAM) |
| $N_z$ | 3 | $T_\varepsilon$ | 300 msec | $R_O$ | 400 MB (RAM) |
| $M_z$ | 1-500 | $T_{int}$ | 10 msec | $R_\varepsilon$ | 10240 MB (RAM) |
| $\beta$ | M+1 | $T_{ext}$ | 15 msec | | |

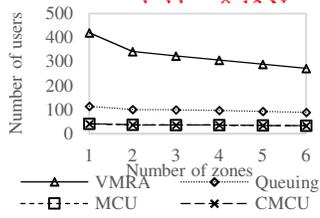
Fig. 4. Maximum users that can be served in a zone

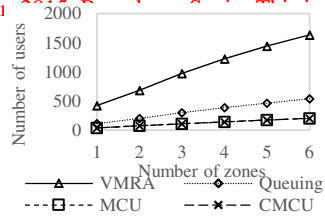
Fig. 5. Total number of users that can be served across all zones

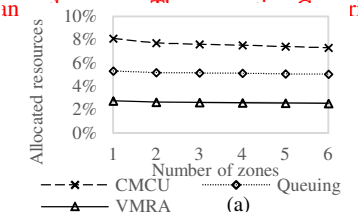
Fig. 6. (a) Average, (b) maximum allocated resources in a data center in Meet-By-All scenario

allocation figures. Other baselines allocate resources as needed. VMRA has better results compared to the other baselines, in both average and maximum cases in this scenario. This is because, the maximum number of users in this scenario is equal to the number of users that MCU can support and just one computation entity is enough to serve them. However, the queuing model, based on our simulation settings, always uses 3 servers to accommodate users. Whereas, VMRA uses 2 VMs to accommodate the same number of users, which leads to the allocation of fewer resources, compared to the queuing model and more resources, compared to MCU and CMCU. However, because the total available resources in VMRA are more than those of MCU and CMCU, the allocated resource percent is lower in comparison to both.

*b) Video Mixing Response Time: Meet-By-All Scenario*

The average video mixing response time for the Meet-By-All scenario is shown in Fig. 7. As it can be seen, queuing model shows better video mixing response time than VMRA. This is because the objective of our model is maximizing resource utilization while respecting QoS. Intuitively, for lower response time, we should allocate more resources; however, this is in contradiction to our objective. So, in VMRA, as long as video mixing response time is lower than QoS threshold, it does not reduce video mixing response time. On the other hand, MCU and CMCU models have more video mixing response time, in comparison to VMRA. This is directly attributed to the centralized architecture of these models. Interestingly, the video mixing response time for MCU and CMCU are the same. It shows cloud has effect only on the amount of allocated resources in CMCU and not on the video mixing response time.

*c) Resource Utilization: Meet-By-Some Scenario*

Recall, our model can serve the maximum number of users, as shown in Fig. 5. Hence, in this scenario, we have as much users as VMRA can serve. As depicted in Fig. 8(a) and 8(b), the resource allocation of the queuing model performs better compared to VMRA. This is because, VMRA will add more resources to accommodate as many users as possible, within the QoS threshold, while queuing model serves requests by leveraging fixed number of servers.

*d) Video Mixing Response Time: Meet-By-Some Scenario*

Previous results show queuing model allocates lower amount of resources in Meet-By-Some scenario compared to VMRA. However, this model is not suitable for video mixing as a service after comparing the corresponding video mixing response time. This is because queuing model sacrifices QoS to serve the same number of users, compared to VMRA. As shown in Fig. 9, if we choose resource allocation based on queuing model for video mixing as a service in cloud we have a high violation in terms of QoS. Based on our simulation results, if we serve as much users as VMRA can support using the queuing resource allocation model, QoS will be sacrificed between 66% and 72%. The same holds true when comparing with CMCU. In fact, VMRA allocates more resources, compared to queuing model and CMCU, to satisfy QoS for more users.

It is important to note that Fig. 6 and Fig. 8 reveal that to accommodate larger number of users for video mixing, it is desirable to have more data centers with fewer resources. Furthermore, as evident from the results, our novel VMRA addresses the specific needs of video mixing as a service, which cannot be handled by generic cloud-based resource allocation models.

## VI. CONCLUSION

We propose a novel and scalable, with respect to number of users, Video Mixing Resource Allocation (VMRA) model for multimedia conferencing applications. We optimally utilize leased resources and dynamically allocate and release resources for changing number of users, while meeting QoS end-to-end delay for video mixing response time. We model VMRA as an optimization problem and design a heuristic for large-scale scenarios. Simulation results show that our VMRA model outperforms other resource allocation techniques for video mixing because it considered both resource efficiency and video mixing QoS requirements. Future work includes, extending the VMRA model to account for VM instantiating time and modeling the VMRA problem from the perspective of the multimedia conferencing provider to have video mixing as a service with minimum cost.

### ACKNOWLEDGMENT

This work is supported in part by the NSERC SAVI Research Network, the FQRNT Team Program, and the NSERC Discovery grant program.

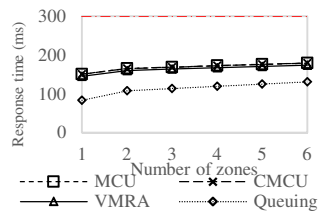
Fig. 7. Average video mixing response time in Meet-By-All scenario

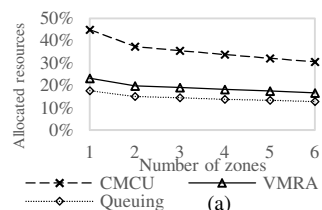
Fig. 8. (a) Average, (b) maximum allocated resources in a data center in Meet-By-Some scenario

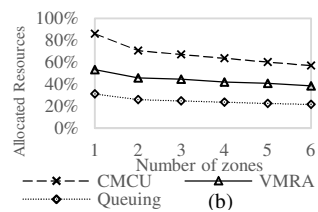

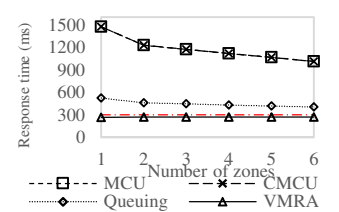
Fig. 9. Average video mixing response time in Meet-By-Some scenario